# Ultrafast modulation of covalency in GeTe driven by a ferroelectric soft mode

Bulat Burganov[1], Vladimir Ovuka[1], Matteo Savoini[1], Helmut Berger[2], J. Hugo Dil[2,3], Juraj Krempasky[3], Steven L. Johnson[1,4]

[1]Institute for Quantum Electronics, Physics Department, ETH Zurich, 8093 Zurich, Switzerland.

[2]Institute of Physics, Ecole Polytechnique Federale de Lausanne, 1015 Lausanne, Switzerland

[3]Photon Science Division, Paul Scherrer Institut, 5232 Villigen, Switzerland

[4]SwissFEL, Paul Scherrer Institute, 5232 Villigen-PSI, Switzerland.

**Abstract**

The general idea of using ultrashort light pulses to control ferroic order parameters has recently attracted attention as a means to achieve control over material properties on unprecedented time scales. Much of the challenge in such work is in understanding the mechanisms by which this control can be achieved, and in particular how observables can be connected to structural and electronic properties. Here we report on a combination of experimental and computational methods to study the electronic structure of the semiconducting ferroelectric GeTe when driven out of equilibrium by absorption of a femtosecond pulse of light. We observe coherent modulations of second harmonic generations on the order of 50%, which we attribute to a combination of atomic and electronic structure changes due to a coherently excited soft mode. Comparison of the observed experimental data with model calculations indicates that this effect is predominantly due to an ultrafast modulation of the covalency of the bonding between Ge and Te ions. This stands in contrast to previously held assumptions in other systems, indicating that care should be exercised in using indirect measurements of electronic structure to make strong conclusions about the magnitude of nuclear motions.

**Introduction**

GeTe is a bulk ferroelectric semiconductor with the simplest possible structure, consisting of only two atoms in a rhombohedral unit cell. In addition to its properties as a ferroelectric it has found applications as a thermoelectric, phase change, and spintronic material. These functional properties are rooted in electronic degrees of freedom characterized by anomalous Born charges, strong electron-phonon and spin-orbit couplings and unconventional bonding mechanism. One key property that is strongly connected to all of these phenomena is the breaking of inversion symmetry at room temperature [1–5].

In general, the inversion symmetry breaking (ISB) in a material can be a combination of the structural and electronic components. Second harmonic generation (SHG) is a convenient way to probe the ISB [6–11]. In cases of purely structural ISB, a simple linear relationship between the second order susceptibility $\chi_2$ and the structural order parameter (OP) is usually assumed [12]. However, when the ISB has a significant electronic component (e.g. in electronic ferroelectrics due to spin, orbital, or charge order, or dynamically in superconductors), the relationship between $\chi_2$ and the underlying electronic order parameter is, in general, more difficult to ascertain.

Here we use SHG to probe the ferroelectric state in GeTe in a dynamic setting, where it becomes possible to separate to some extent the structural and electronic components of the ISB. We use displacive excitation of a coherent phonon (DECP) induced by the absorption of a femtosecond laser "pump" pulse to modulate the structural ISB, and observe the dynamics of the SHG as a way to measure $\chi_2$. We find that the second harmonic susceptibility is strongly sensitive to the electronic degrees of freedom; for example, it is directly affected by the reoccupation of electronic states caused by the pump pulse. Crucially, we find that the dependence of $\chi_2$ on the ionic displacement of the polar soft mode is strongly nonlinear, even for very small displacements. This is in contrast to more conventional oxide ferroelectrics. We show that this nonlinearity originates from the charge transfer between the ions and the shorter Ge-Te bond along the polar direction, a process similar to one encountered in electronic ferroelectrics. Finally, we discuss how this charge transfer can be considered as a dynamic modulation of the bonding character in GeTe.

The space group of GeTe lattice is $R3m$, and its low-temperature crystal structure in the ferroelectric phase can be derived from the face-centered cubic NaCl lattice by a small rhombohedral distortion (see inset of Fig. 1(a)). The inversion symmetry is broken by a static displacement of the ions along the [111] body diagonal of the unit cell. The three optical modes are all simultaneously Raman and IR active. One mode is of $A_1$ symmetry where the Ge and Te ions move in opposite directions along the [111] direction. The remaining two modes are of $E_1$ symmetry and consist of displacements of the ions in directions orthogonal to [111].

**Methods and results**

The sample is a single crystal of GeTe grown from the vapour phase by dissociative sublimation and by chemical transport with Germanium(IV) iodide (GeI$_4$) as transport agents in sealed ampules, with one surface oriented and polished to the (010) plane with an accuracy of 10°. The dynamics of both the linear optical properties and the second harmonic generation are measured in a reflection geometry. An amplified 250 kHz Ti:sapphire laser system delivers 45-fs-long 800 nm pulses used for both the pump and the probe. The pump pulse excites the sample, with an incidence normal to the sample surface and a spot size of 70 μm. The incident fluence is set to a value between 3 and 9 mJ/cm$^2$. The probe pulse reflects from the sample at an angle of 6° to surface normal and has a spot size of 32 μm on the surface. The probe is then sent through a polarizer and then focused onto either a photodiode (for linear reflectivity measurements) or a photomultiplier tube (for second harmonic generation). For the second harmonic generation measurements a bandpass filter (400 nm center wavelength, 40 nm bandwidth) is placed in front of the detector. The pump and probe pulses are cross-polarized to reduce pump scatter on the detector, and sample azimuth is oriented to maximize the measured second harmonic signal. The probe pulse energy was set to 9 nJ was for the SHG measurements, resulting in an incident fluence of 1.1 mJ/cm$^2$. For the reflectivity dynamics, we reduce the probe fluence to 0.05 mJ/cm$^2$. All measurements were performed at 295 K, much lower than the ferroelectric Curie temperature of 705 K [13].

Similar to what has been reported elsewhere [14], we observe oscillations of the reflectivity with a period of approximately 0.27 ps (Fig. 1(a)). The frequency of these oscillations corresponds with that of the A$_1$ polar mode, which is displacively excited by the sudden change in electronic state occupation caused by the pump pulse [15-16]. Assuming a locally harmonic potential, we can model these dynamics as damped oscillations of the polar mode according to

$$Q_P(t) = q_1 + (q_0 - q_1) \cos \omega t \, e^{-\gamma t}. \tag{1}$$

Here, $Q_P$ describes the relative displacement of the Ge and Te ions along the [111] direction, where a value of zero indicates that the Ge ion is in the center of the unit cell as drawn in the inset of Fig. 1(a). Concretely, $Q_P = d_{Ge-Te} - d_0$, where $d_{Ge-Te}$ is the distance along [111] between one particular pair of Ge and Te ions, and $d_0 = |\vec{a}_1 + \vec{a}_2 + \vec{a}_3|/2$ is the distance along this direction from the Te ion to the center of the unit cell. The parameter $q_0$ denotes the equilibrium value of $Q_P$ before laser excitation, while $q_1$ is a new quasiequilibrium value in the electronically excited state. The parameters $\omega$ and $\gamma$ are the frequency and damping of the A$_1$ vibrational mode, the frequency depends on the pump power density and is in the range 3.8 to 3.0 THz [14].

To estimate the values of $q_0$ and $q_1$, we use the DFT calculations of the reconstructed potential energy surfaces (Fig. 1(b)). Our calculation assumes that electrons thermalize instantaneously to a single chemical potential after the absorption of the pump pulse, similar to previously reported work on Bi [17]. The energy at an elevated temperature is calculated from $E_{tot}(Q_P, T_e) = E_{tot}(Q_P, 0) + \Delta E(Q_P)$, where $E_{tot}(Q_P, 0)$ is the self-consistent total energy of the electronic ground state for a fixed value of the polar distortion $Q_P$, and $\Delta E$ is calculated by populating ground-state Kohn-Sham states according to a single Fermi-Dirac distribution with temperature $T_e$. In this approach, the energies of the Kohn-Sham states are interpreted as single-electron excitation energies. These calculations were done with the lattice constants fixed to their experimental values at 295 K: $a$ = 5.985 Å, $\alpha$ = 88.17° [13].

The potential energy surfaces on which the ions move are characterized by constant electronic entropy rather than temperature [18] and are shown in Fig. 1(b) (inset). The zero-entropy (ground state) potential minimum is found to be $q_0$ = 28 pm, which compares well with the experimental value from neutron scattering of 26.5 pm at 295 K [13], and 31 pm from X-ray data for thin films [19]. An absorbed energy density of $\Delta E$ = 0.22 eV/u.c. increases the entropy to $S_e$ = 0.85 $k_B$/u.c. ($T_e$ = 4522 K) and shifts the potential minimum to $q_1$ = 20.4 pm (by 27% of the initial value). Up to this excitation level, the relationship between $q_1$ and $\Delta E$ is approximately linear (Fig. 1(b)). The phase transition to a high symmetry state ($q$ = 0) happens at $\Delta E_c$ = 0.46 eV/u.c. ($S_e$ = 1.37 $k_B$/u.c. and $T_e$ = 6044 K).

We now turn to our measurements of the changes in the second harmonic generation in response to the pump pulse. Fig 2(a) shows a summary of these measurements for a range of pump fluences. The observed transient SHG response can be described as a sequence of three phases: 1) an initial abrupt suppression of signal followed by oscillations around a new equilibrium, 2) a rapid partial recovery within 2.5-4 ps, and 3) a slower exponential recovery after 4 ps.

We interpret the initial abrupt drop of the signal as a direct consequence of the redistribution of electronic states after excitation. The oscillations that follow have the same periodicity as those seen in the reflectivity, and thus are likely also due to the coherent excitation of the $A_1$ polar mode. We tentatively assign the recovery during the phase 2 with the cooling of the electronic subsystem as it thermalizes with the lattice. The change of slope at 2.5-4 ps would then correspond to the time $\tau^*$ when the lattice and electronic temperatures equilibrate. The slow SHG changes after $\tau^*$ are set by the time scale for the excited region of the sample to cool via diffusive heat transport into the bulk. At these late times we also see slow oscillations superimposed on the recovery, with a period of about 20 ps. These are not seen in the reflectivity data. We assign these to acoustic strain waves, which have been observed in other systems to have a much stronger effect on SHG than on properties derived from solely linear optical properties [20].

We focus first on the oscillations observed during the first 1.5 ps after the pump pulse. In the analysis of SHG from ferroelectrics it is usually assumed that $\chi_2$ is proportional to the ferroelectric polarization $P$ [9, 21, 22]; the same result follows from a thermodynamic description of ferroelectrics [23]. Furthermore, it is commonly assumed that the ferroelectric polarization is proportional to the ionic displacement, or polar coordinate $Q_P$: $P = ZQ_P$, where $Z$ is the effective charge [24, 25]. Thus, as an initial model to describe time-resolved SHG response, we presume that the second order susceptibility is proportional to the displacement: $\chi_2/\chi_2^0 \propto Q_P(t)/q_0$, where $\chi_2^0$ is the SH susceptibility at $t < t_0$.

The observed oscillations are superimposed on a background that is nearly constant or even decreasing during the first 0.5 ps after $t_0$ and then starts increasing after ~1 ps. This non-exponential behaviour has been observed in Bi [26] and it was linked to the changes in equilibrium atomic positions within the first picosecond during thermal equilibration of the electrons and holes, which is not captured in our model since we assume instantaneous electronic thermalization. We assume that the remaining dynamics in the SHG are due to changes in $\chi_2$ caused by the $A_1$ optical phonon. We thus model the short-time changes in SHG after the pump excitation at $t = 0$ as

$$\frac{I_{SHG}}{I_{SHG,0}} = \left(\frac{Q_P(t)}{q_0}\right)^2 (1 - \Delta + at + bt^2) \qquad (2)$$

where $Q_P(t)$ follows Eq. (1), $\Delta$ is a parameter reflecting the immediate change in SHG efficiency from the pump-driven electronic state redistribution, and the parameters a and b are phenomenological parameters that take into account possible effects of electronic state redistribution on a time scale of 100's of femtoseconds. The fluence dependences of the fitting parameters $q_1$ and $\Delta$ are shown in Fig. 2(b). At $t < 0.8$ ps, the term $at + bt^2$ is much smaller than $\Delta$ and thus the use of this polynomial background does not strongly affect the extracted values of $q_1$ and $\Delta$. A good model should correctly reproduce three aspects of the data for each fluence: oscillation frequency, the step size at $t = t_0$ and the oscillation amplitude. Overall the fit quality in Fig. 2(a) is acceptable; however, the dependence of the fit parameters on fluence in Fig. 2(b) is unexpected for several reasons. First, the $q_1$ dependence on fluence is not linear as would be expected from the DFT calculations for this range of displacement values (Fig. 1(b)). Second, the SHG signal suppression due to electronic excitation $\Delta$ is a *decreasing* function of fluence, which seems contrary to the expectation that larger drops should be seen at higher excitation levels. This last observation can be seen directly in the data by comparing the ratio of the step size to the oscillation amplitude as a function of fluence: one sees that this ratio decreases strongly with increasing excitation.

Independent of the DFT calculations, the observed dependence of $q_1$ on fluence $F$ is also inconsistent with a simple, phenomenological quartic double-well model of the potential energy surface $U = (1 - F/F_c)z^2 + z^4/2$ that has been previously used to describe dynamics of structural order parameters in an excited electronic state near a second order phase transition [27]. This parameterization implies a quasi-equilibrium structural coordinate $z_1 = q_1/q_0 = \sqrt{1 - F/F_c}$, and thus predicts a nonlinear dependence on fluence that at first glance is at least qualitatively consistent with our data. In Fig. 2(b) we plot $z_1$ versus $F$ for various values of $F_c$ as dashed lines, and compare this to the results of our fit. There is still significant disagreement, as the observed dependence of $z_1$ on the excitation fluence is much stronger than even this model predicts. A more convincing model of the SHG dynamics over this time range appears to require a more general model that does not assume the second order susceptibility is simply proportional to the polar structural distortion.

In order to construct such a model, we first posit a generalized form of the dependence of the second-order susceptibility on the coordinate $Q_P$ and the change in electronic energy $\Delta E$:

$$\chi_2(Q_P, \Delta E) = \chi_2^0 f(Q_P/q_0) g(\Delta E), \qquad (3)$$

where in principle the functions $f$ and $g$ are arbitrary functions of the normalized polar coordinate and the change in electronic energy, respectively. Our treatment in equation (2) is equivalent to assuming that $f(z) = z = Q_P/q_0$. Since symmetry considerations imply that $f(z)$ must be odd, the next order of approximation is $f(z) = (1 + c)z - cz^3$, where $c$ is some constant and we have applied the constraint that $f(1) = 1$. As a next step, we must construct a form for $g(\Delta E)$, which gives the dependence of the second order susceptibility on the change in electronic energy density $\Delta E$. For guidance, we look at the dependence of the second harmonic signal immediately after the pump excitation, as a function of fluence (Appendix A).

Over the range of fluences measured between 3 and 9 mJ/cm² we see a nearly linear dependence of $g$ on incident fluence. Projecting this approximately linear dependence to zero fluence leads, however, to a value of $g(F=0) = 0.94$ which disagrees significantly with the trivial observation that zero fluence leads to no change, or $g(0) = 1$. We thus model this dependence as $g(\Delta E) = g_1 e^{-\Delta E/\Delta E_1} + (1-g_1)e^{-\Delta E/\Delta E_2}$, where $g_1$, $\Delta E_1$ and $\Delta E_2$ are model parameters.

We model thermal and structural dynamics in the sample following the absorption of the pump pulse at each fluence value using the two-temperature approach and DFT-derived potential energy surfaces (Appendix B). Then the local dynamics of the second order susceptibility at depth $z$ are governed by equation (3). The volume of the sample probed by SHG is confined by the probe spot size and depth given by the coherence length for SHG in reflection $l_{coh} = \pi/|k_{2\omega} + 2k_\omega| \approx 19$ nm. The detected SHG intensity is found in the first Born approximation by integrating contributions from all depths:

$$I_{SHG}(t)/I_{SHG}(0) = \left| \int_0^\infty \chi_2(z,t)/\chi_2^0 \exp(i\Delta k z)\, dz \right|^2, \qquad (4)$$

where $\Delta k = k_{2\omega} + 2k_\omega$. The effect due to finite propagation time through the probed volume is neglected as it is much smaller compared to the time resolution and the relevant time scales of the measurement. Finally, the simulated dependences $I_{SHG}(t)/I_{SHG}(0)$ are fitted to the data and plotted on Fig 3(a). The fit parameters and their fitted values are $c = 0.3\pm0.1$, $g_1 = 0.05\pm0.02$, $\Delta E_1 = 0.04\pm0.02$ eV/u.c., $\Delta E_2 = 9\pm2$ eV/u.c., and the corresponding dependences $f(z)$ and $g(\Delta E)$ are shown in Fig. 3(b,c). Fluence values in the simulation are rescaled by a factor of 0.65 from the experimental values to produce the best match with the data. The distribution of $\Delta E$ and polar coordinate values within 20 nm below surface are shown as histograms in Fig. 3(b, c).

Finally, we use a two-temperature model with variable electron-phonon coupling constant $G_0$ to match the equilibration time $\tau^*$ with the experimental values at different fluences (Fig 2(c)). The values of $G_0$ agree by order of magnitude with the ones calculated from the first principles for hole concentration of $2.1 \times 10^{21}$ cm$^{-3}$ [28].

**Discussion**

We focus here on the nonlinear dependence of $\chi_2$ on the polar coordinate. We first compare the fitted dependence $\chi_2(Q_P)$ to the dependence of polarization on $Q_P$. In Fig. 4(a) we show the polarization calculated from the DFT-derived electronic structure using the Berry phase method [29, 30] for a series of polar distortion values while keeping the lattice parameters fixed to the ferroelectric structure at 295 K. The calculated polarization displays a nonlinear behaviour. We fit a cubic polynomial to $P(Q_P)$ and estimate the dynamic charge $Z^* = \partial P/\partial Q_P$ [25] as a fraction of the static one: $(Z^*/Z)_{theo,P} = \partial P/\partial Q_P * Q_P/P = 0.58$. Next, we can estimate an analogous quantity from $\chi_2(Q_P)$ dependence (assuming $\chi_2 \propto P$): $(Z^*/Z)_{exp,SHG} = \partial f/\partial x|_{x=1} = q_0/\chi_2^0 * \partial \chi_2/\partial Q_P = 0.4\pm0.2$.

Dynamical contributions to the effective charge $Z^* = \partial P/\partial Q_P = Z + Q_P \partial Z/\partial Q_P$ indicate that the effective charge changes rapidly with the bond length or $Q_P$. This change appears to be unusually large in GeTe, originating from an additional transfer of charge not simply proportional to the ionic motion $Q_P$ [25]. For comparison, we also show the calculated $P(Q_P)$ dependence for $BaTiO_3$ and $PbTiO_3$, in which the dynamical contribution is very small. The possibility of such anomalous behaviour of the effective charge in GeTe was theoretically proposed previously [31]. Here we find the first experimental evidence and a quantitative estimate of such a dynamic contribution.

Next, we discuss the qualitative mechanism by which $Q_P$ can influence the SHG at a specific optical wavelength. In GeTe, direct electronic transitions at 1.55 eV are available near the L and Z points of the Brillouin zone (Fig. 4(b)); hence, the SHG intensity can be particularly sensitive to the dynamic charge contribution associated with these transitions. In particular, the second-harmonic susceptibility is sensitive to the change of the dipole moment between the states involved in the transition [32–35]. We use a proxy for the dipole moments by calculating a displacement of the real-space density center of mass (COM) of the Kohn-Sham eigenstates with change of the polar coordinate from 0 to $q_0$. For the state $\psi_{i,k}$ in band $i$ and momentum $k$, the state COM is found as $r_{cm,i,k} = 1/v_{u.c.} \int r|\psi_{i,k}(r)|^2 dv$. Each band in Fig. 4(b) is color-coded by COM displacement along [111], $\Delta r_{cm,i,k} = r_{cm,i,k}(Q) - r_{cm,i,k}(Q = 0)$. Valence band state $V_3$ and conduction band state $C_2$ at $k = Z$ have the largest COM displacements, and, hence are likely to be the most important for the inversion symmetry breaking by the electronic degrees of freedom. Next, we show the $Q_P$ dependence of the COM for the states near the Fermi level at $k = Z$ in Fig. 4(d). The dependence is found to be strongly nonlinear. Charge distribution of states $V_3$ and $C_2$ at $k = Z$ for the centrosymmetric ($Q_P = 0$) and ferroelectrically distorted ($Q_P = q_0$) structures is shown in Fig. 4(c). The electron density of state $V_3$ surrounding the Te ion in the centrosymmetric phase is transferred to the shorter Ge-Te bond along the

[111] direction in the distorted structure. This charge transfer is incommensurate with the ion motion, and can be quantified by the dynamic contribution to the effective charge mentioned above. It is thus possible to dynamically control this charge transfer via controlling the $Q_P$ coordinate.

This in turn provides a fundamental insight into the interaction of the electronic system with photons of 1.55 eV energy. The electronic excitations near the $k = Z$ point involving the states $V_3$ and $C_2$ contribute strongly to the SHG process, since the change of the dipole moment on excitation is the largest. We expect the same transitions to be the most effective in reconstructing the potential surface by the pump pulse, since the states below and above the gap favour opposite ferroelectric polarizations. A full *ab-initio* modelling of the time-dependent SHG process could provide further quantitative insight but is outside the scope of this work.

The unusually large value of the dynamic charge transfer can actually significantly affect the bonding character in GeTe. The nature of chemical bonding in GeTe is important for its functional properties, including thermoelectric [38], phase-change [39], and ferroelectric property [31, 40]. The concept of metavalent bonding has recently been proposed for a group of chalcogenides that combine properties of covalent and ionic bonding, and it has been shown that covalency can be affected by the structure [41, 38, 39]. We plot a quantitative measure of covalency, the number of electrons shared between Ge and Te on the short bond, as a function of ferroelectric distortion, from cubic ($u = 0.25$) to the fully distorted ($u = 0.2365$) in Fig. 4(e) (reproduced from [42]). The number of shared electrons varies between 0.9 and 1.24 within these limits. In our experiment, $Q_P$ is modulated in the range $(0.55q_0, q_0)$, and hence the number of shared electrons is modulated approximately in the range (1.09, 1.24) electrons. In contrast, the ionicity, measured by the number of electrons transferred from Ge to Te is nearly independent of the distortion [42].

We note that the dynamic charge transfer along the polar bond occurs in oxide ferroelectrics [43, 44], in which the covalent bonding was found to be essential to stabilize ferroelectricity [44–47]. The charge transfer in GeTe, that we found evidence of, has relatively very large magnitude. A large charge transfer can be at origin of electronic ferroelectricity in molecular crystals [48–51], where the total polarization is driven by the electronic contribution and can be even antiparallel to the ionic displacement [48]. High electronic polarizability, large electron-phonon coupling and anomalous Born charges, all of which are present in GeTe, have been identified as characteristics of electronic ferroelectrics [52]. Thus an investigation of a possible electronic origin of GeTe ferroelectricity seems to be potentially fruitful. Finally, the dynamic charge transfer found here can help explain the observed strong magnetoelectric coupling in $Ge_{1-x}Mn_xTe$ [5, 53] and provide a pathway to design novel multiferroic materials.

In conclusion, we report on the dynamics of the second harmonic generation in GeTe following the excitation by an IR pump. The SHG intensity was found to be modulated by over 50%, in contrast to a few percent modulations of reflectivity. By taking advantage of the simple crystal structure and knowledge of the structural dynamics, we demonstrated the anomalous behaviour of the SH susceptibility indicative of a large dynamic transfer of charge incommensurate with the ionic motion. Our work opens up new avenues for ultrafast control of functionality by light through manipulation of the degree of covalency in chalcogenides.

ACKNOWLEDGEMENT

This work was supported by the NCCR Molecular Ultrafast Science and Technology (NCCR MUST), a research instrument of the Swiss National Science Foundation (SNSF), and SNSF grant 200021_169698. Parts of the work were also supported by SNSF project grants 200021_169698 and 200020_192337.

## Appendix A. Model of time-dependent second harmonic susceptibility

We model the polar mode dynamics using Eq. (1) and the susceptibility for second harmonic generation $\chi_2$ using Eq. (3) in order to reproduce the observed coherent oscillations at $t \lesssim 1.2$ ps (Fig. 5(a)). The dependence on polar coordinate is given by $f(z) = (1 + c)z - cz^3$, where $z = Q(t)/q_0$. $q_0$ and $q_1 = z_1 q_0$ are the minima of the potential energy $V(Q_P)$ immediately before and after the excitation, respectively. Motivated by DFT calculations (Fig. 1(b)), we assume a linear dependence $z_1 = 1 - aF$, where $a$ is a scalar.

The argument of function $g$ is local excitation energy density $\Delta E$, which is proportional to pump fluence $F$, assuming a homogeneous excitation within length scale $l$: $\Delta E = (1 - R)\, F/l$, where $R$ is surface reflectivity. The dependence of $g$ on fluence can be derived from the knowledge of $f(z_1)$ and SHG intensity at zeros of oscillation $t'$ ($\cos \omega t' = 0$), since $\chi_2(t')/\chi_2^0 = \sqrt{I_{SHG}(t')/I_{SHG}^0} = f(z_1)g(F)$. We plot $\chi_2(t')/\chi_2^0$ as a function of fluence on Fig. 5(b), where $t'$ are the first zeros of the oscillations.

Next, we fit the calculated time-dependent SHG traces using $a$ and $c$ defined above as fit parameters (Fig. 5(a)). Fig. 5(b) shows the fluence dependences of $z_1 = 1 - aF$, $f(z_1) = (1 + c)z_1 - cz_1^3$ and $g(F) = \chi_2(t')/\chi_2^0/f(z_1)$, corresponding to the best fit: $a = 0.028$ (mJ$^{-1}$cm$^2$) and $c = 0.45$. The values $g(F)$ display a linear dependence, however, to satisfy the constraint $g(0) = 1$, we fit this dependence with a sum of exponentials:

$$\chi_2(Q_P = q_0, F)/\chi_2^0 = \sum b_k\, e^{-\alpha_k F},$$

with $b_1 = 0.057$, $b_2 = 1 - b_1$, $\alpha_1 = 1.97$ (mJ/cm$^2$)$^{-1}$, $\alpha_2 = 0.009$ (mJ/cm$^2$)$^{-1}$.

Below we discuss a possible physical interpretation for the dependence of $\chi_2$ on $\Delta E$ as a sum of exponentials. The interaction cross section of the IR pulse with electrons depends on the availability of states above the gap satisfying the resonance condition: $E_f = E_i + h\nu$, and thus, in general, varies for different electronic states. For simplicity, we consider two electronic subsystems, each with population $n_k$ in the valence band. If during time $dt$ a portion of the energy $\delta E = I dt$ supplied by the pump pulse is absorbed, then the populations change by $dn_k = -\alpha_k n_k I dt = -\alpha_k n_k \delta E$, where $\alpha_k$ is used to quantify the strength of the interaction for each of the populations. We only consider the dependence on the number of valence electrons $n_k$ (below the gap) and assume that once excited, they quickly thermalize above the gap. The populations after absorption of the full pulse energy $\Delta E$ are then found by integration: $n_k = n_{k,0} e^{-\alpha_k \Delta E}$. If we additionally assume that the contribution of each subsystem to the SH susceptibility is proportional to its population, $\chi_{2,k} = \beta_k n_k$, then the total susceptibility is $\chi_2 = \sum \beta_k n_{k,0} e^{-\alpha_k \Delta E}$. The saturating behaviour in Fig. 3(c) is an indication of bleaching of the electrons in one of the two subsystems by the pump pulse. The fitted values $\alpha_k = 1/\Delta E_k$ indicate strong interaction of the IR pulse with the subsystem 1 ($\Delta E_1 = 0.04 \pm 0.02$ eV/u.c.) that contributes a fraction $g_1 = \beta_k n_{k,0} = 0.05 \pm 0.02$ to the second harmonic susceptibility, and a relatively weak interaction with subsystem 2 ($\Delta E_2 = 9 \pm 2$ eV/u.c.).

The simulation described above motivates using nonlinear dependences of $\chi_2$ on excitation density and polar mode $Q_P$. However, this model has several drawbacks: 1) inhomogeneity of excitation in the sample is not considered, 2) the dependence of $z$ on fluence is assumed to be linear, 3) the electronic and lattice temperatures are constant, and thus time-dependent oscillation frequency is not supported. These disadvantages are addressed with a more complete model based on the two temperature approach and the polar mode dynamics calculated using DFT (Fig. 3(a)).

## Appendix B. Two-temperature modelling

We use the two-temperature approach [54], to model thermal and lattice dynamics in the sample after the absorption of the pump pulse. Due to the small penetration depth of 37 nm of the 800 nm laser pulse compared to the focused pump spot size of about 70 μm x 70 μm, the variation of temperatures is effectively one dimensional, i.e. perpendicular to the sample surface. We solve a system of differential equations [18]:

$$T_e \frac{\partial S_e}{\partial t} = P + \frac{\partial}{\partial z}\left(\kappa_e \frac{\partial T_e}{\partial z}\right) - g_0(T_e - T_l),$$

$$C_l \frac{\partial T_l}{\partial t} = D_l \frac{\partial^2 T_l}{\partial z^2} + g_0(T_e - T_l),$$

where $z$ is the distance from the sample surface, $T_e$ and $T_l$ are electronic and lattice temperatures, $S_e$ – electronic entropy, $P$ is the energy deposited by the laser pulse per unit volume and time, $g_0$ – electron-phonon coupling constant governing the heat transfer from electrons to phonon bath, $\kappa_e$ is electron thermal conductivity, $D_l$ is lattice thermal diffusivity. The source term is given by

$$P(z,t) = \frac{2F(1-R)}{l_p t_w} \sqrt{\frac{\ln 2}{\pi}} \exp\left[-4\ln 2 \frac{t^2}{t_w^2}\right] \exp\left[-\frac{z}{l_p}\right],$$

where $F$ is the incident pump fluence, $R$ – surface reflectivity, $l_p = 37$ nm is penetration depth at a wavelength of 800 nm, $t_w = 45$ fs is the laser pulse FWHM.

The dynamics of the polar mode $Q_P$ are governed by the following equation of motion:

$$\frac{\partial^2 Q_P}{\partial t^2} = -\frac{1}{\mu}\frac{\partial E}{\partial Q_P} - \frac{2}{\tau}\frac{\partial Q_P}{\partial t},$$

where $\mu$ and $\tau$ are the mode effective mass and lifetime, respectively. The energy surfaces $E(Q_P, T_e)$ are calculated from DFT as described in the main text, and are shown in Fig. S4. We also show electronic specific heat, $C_e = \partial E/\partial T_e$, and mode frequency and equilibrium position as a function of electronic entropy $S_e$. The equations are solved numerically in a 200 nm-thick slab dividing the sample into layers with a thickness 2 nm along the interior surface normal.

For the lattice heat capacity we use the theoretical high-temperature value of 3R, which is very close to the measure values at room temperature and above [55, 56]. We ignore the temperature dependence of $C_{lat}$ since it is nearly constant above the Debye temperature in GeTe of 205 K [57]. The experimental values of heat conductivity at room temperature were reported in the range 0.1-4.1 W/mK [58] varying strongly with disorder. We use the value of 3.08 W/mK corresponding to crystalline undoped GeTe [59].

At high electronic temperature, the heat transport is dominated by electrons. We use a theoretically estimated temperature dependence of $\kappa_e$ following the approach outlined in [18], that assumes the dominant electronic scattering through electron-phonon coupling. Thermal transport has no significant effect on the analysis and modeling of oscillations, since the time scale is relatively short. Transport effect do, however, influence the model at later delay times.

# Appendix C. Details of density functional theory calculations

The density functional theory (DFT) computations of the total energy, band structure and Berry-phase polarization were performed with a plane wave basis set, the generalized gradient approximation (GGA-PBESol) for the exchange-correlation functional as implemented in the Quantum Espresso [60], and ultrasoft pseudopotentials [61]. We use 12x12x12 k-point sampling of the Brillouin zone and energy cutoff of 80 Ry for wavefunctions and 400 Ry for the charge density.

The calculations of the ground state energy and the band structure were repeated utilizing the all-electron full-potential linearized augmented plane-wave method implemented in Elk code [62]. The wave-function outputs were used to determine the real-space distribution of Kohn-Sham states for Fig. 4(c).

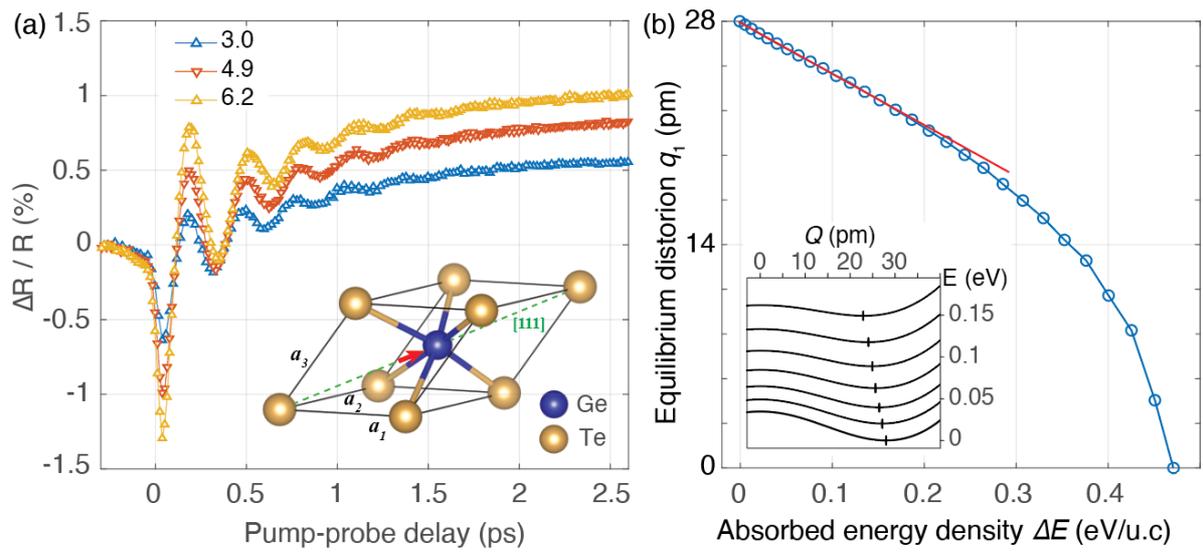

**Figure 1.** (a) Pump-probe reflectivity traces labelled by incident pump fluence (mJ/cm$^2$). Inset: crystal structure of GeTe. (b) Quasi-equilibrium ferroelectric distortion $q_1$ as a function of absorbed energy density $\Delta E$ calculated using DFT. Red line is a guide to the eye. Inset: Potential energy surfaces for the polar mode calculated for the excited states as described in the text.

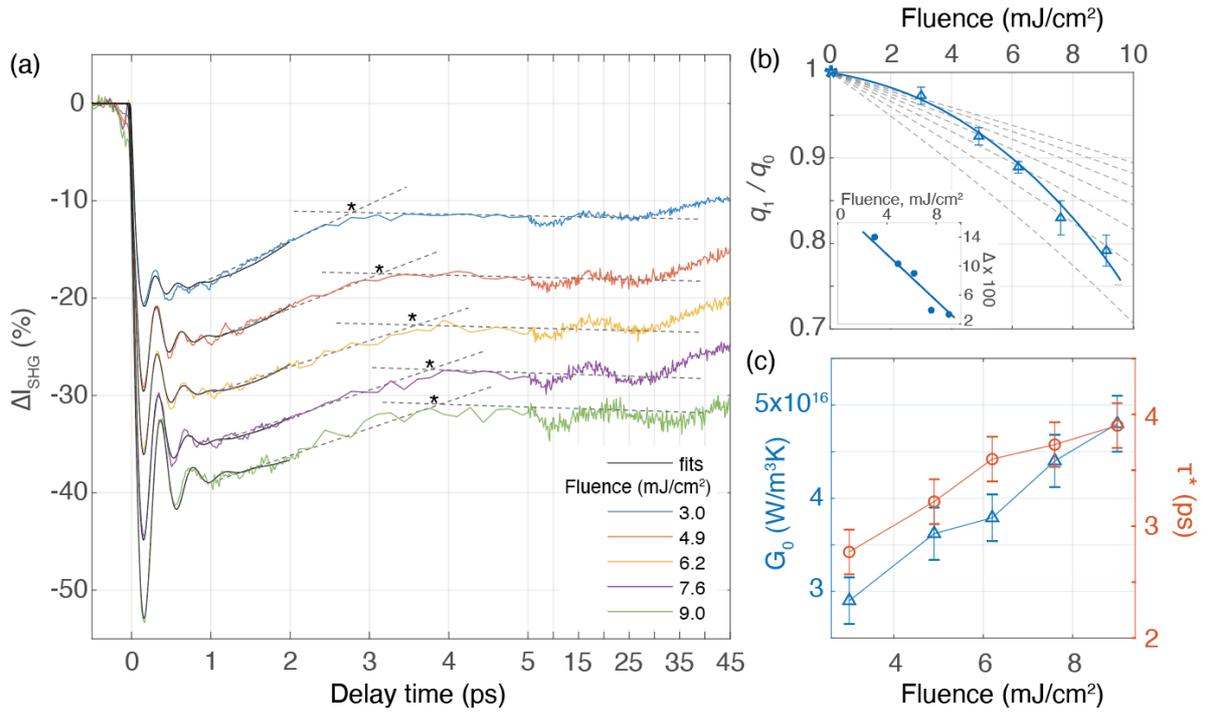

**Figure 2.** (a) SHG dynamics in GeTe and fits using equation (2). Dashed lines are guides to the eye showing the change of slope at $\tau^*$ (see text). (b) Fit parameters describing average ferroelectric distortion $q_1$ and SHG suppression $\Delta$ at $t_0$ due to electronic excitation (inset) as a function of measured incident pump fluence. The dashed lines show dependences $q_1/q_0 = \sqrt{1 - F/F_c}$ for critical fluence $F_c$ between 20 and 50 mJ/cm². (c) Fluence dependence of electron-phonon coupling constant and $\tau^*$.

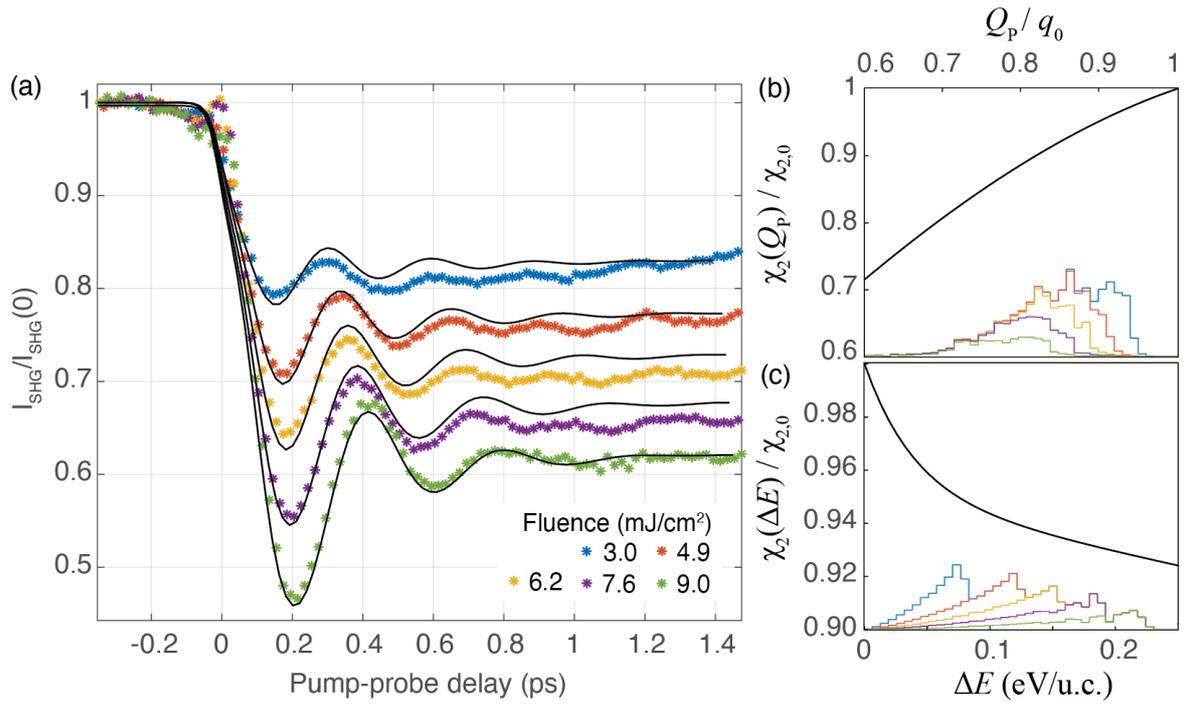

**Figure 3.** (a) Experimental and simulated SHG dynamics. (b) Dependence of SHG susceptibility on the polar coordinate $Q_P$ and (c) on the excitation density. The stacked histograms show the distributions of $Q_P$ and $\Delta E$ values in the simulation for different fluences.

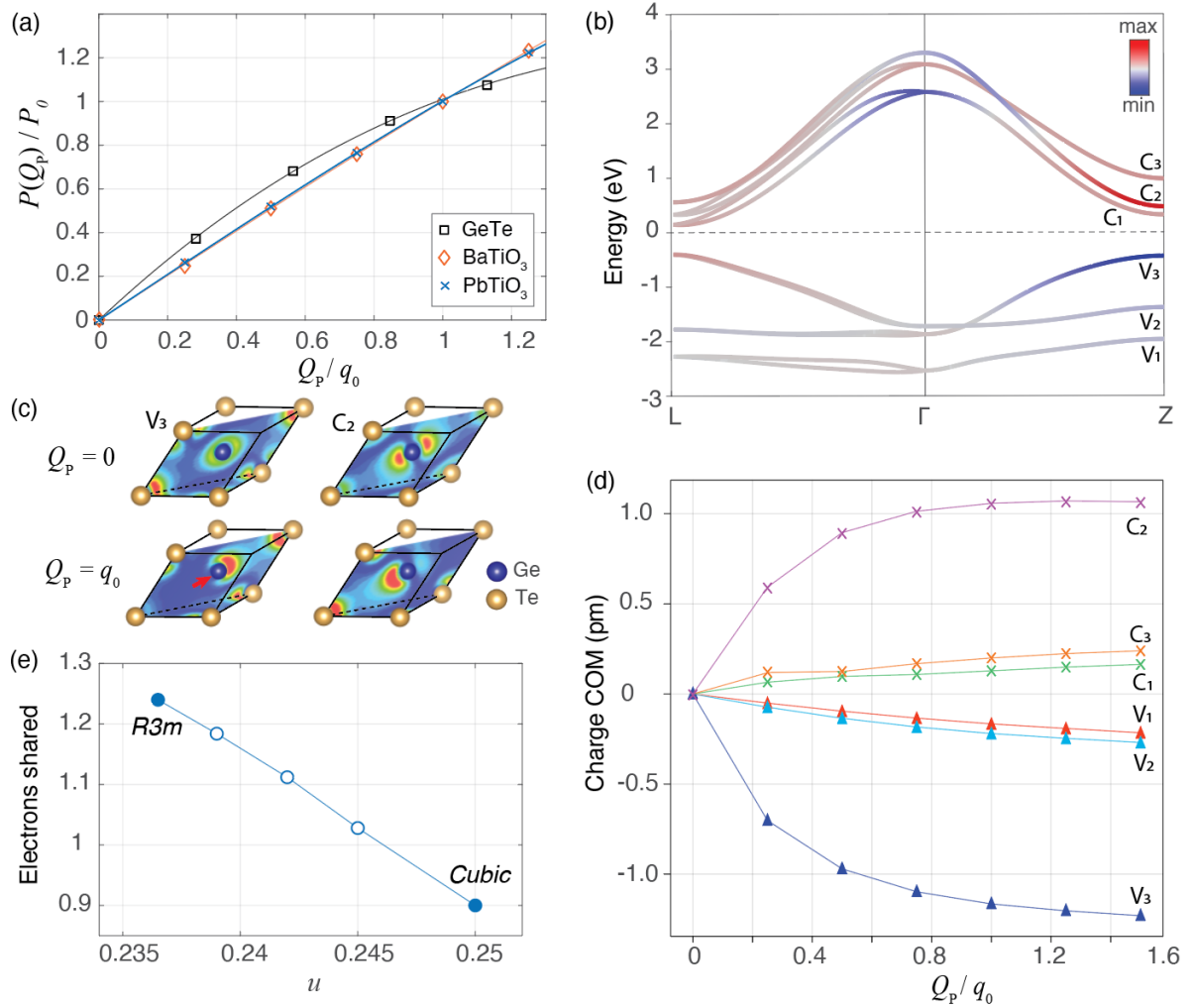

**Figure 4**. (a) Calculated Berry-phase polarization as a function of polar distortion $Q_P$ for GeTe, $BaTiO_3$ and $PbTiO_3$. Lattice parameters are fixed to the values in ferroelectric phase. In $BaTiO_3$ and $PbTiO_3$ the distorted structure ($Q_P = 1$) corresponds to tetragonal phase [36, 37]. Solid lines are cubic polynomial fits. (b) Band structure of GeTe along L-G-Z. Color-coding indicates the state real-space center-of-mass displacement in [111] direction. (c) Real-space density distribution of states at $k$=Z. (d) COM displacement for Kohn-Sham states at $k$=Z as a function of the polar coordinate. (e) Electrons shared between Ge and Te as a function of ferroelectric distortion [42].

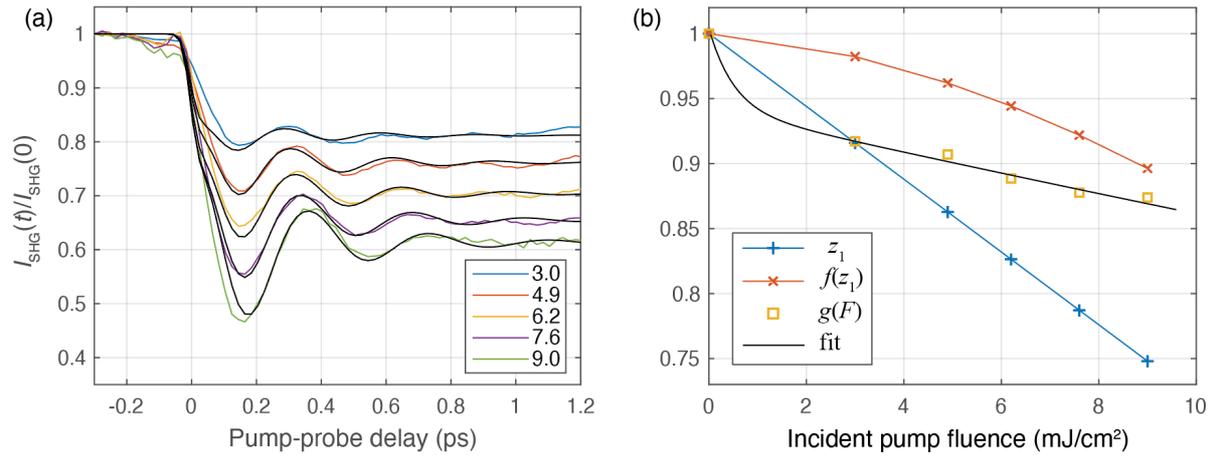

**Figure 5.** (a) Time-resolved SHG dynamics in GeTe. Black solid lines are fits as described in the text. (b) Fluence dependences of polar distortion $z_1$, nonlinearities $f(z_1)$, and $g(F)$.

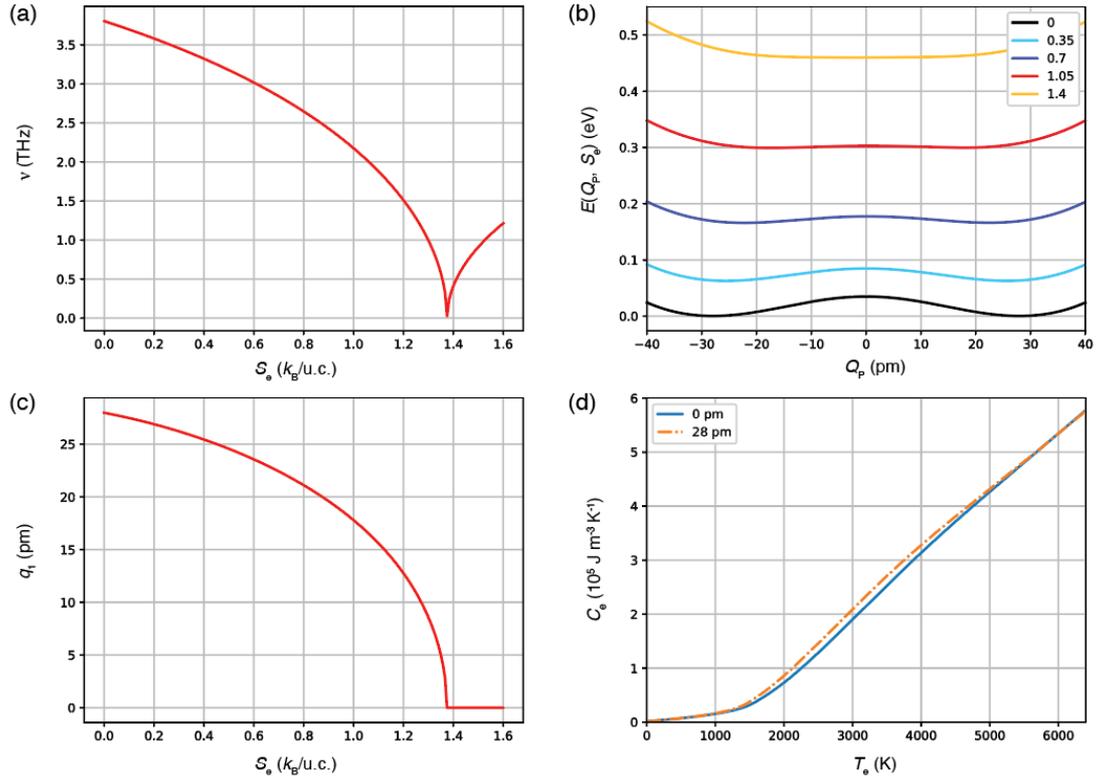

**Figure 6.** (a) Frequency of the polar mode as a function of electronic entropy. (b) Potential energy surfaces for polar mode labelled by the values electronic entropy. (c) Equilibrium displacement (potential minimum) as a function of electronic entropy. (d) Electronic specific heat for centrosymmetric and distorted structures (labelled by the value of $q_1$).